\documentclass[11pt]{article}

\usepackage{graphicx}
\usepackage{lineno}
\usepackage{epsfig}

\newcommand{\BaBarYear}       {05}
\newcommand{\BaBarNumber}     {012}
\newcommand{\SLACPubNumber} {11303}
\newcommand{\BaBarType}     {CONF}

\input pubboard/babarsym

\long\def\inst#1{\par\nobreak\kern 4pt\nobreak
    {\it #1}\par\vskip 10pt plus 3pt minus 3pt}

\begin{document}
{\pagestyle{empty}

\begin{flushright}
\babar-\BaBarType-\BaBarYear/\BaBarNumber \\
SLAC-PUB-\SLACPubNumber \\
\today \\[.7in]
\end{flushright}

\begin{center}
\Large \bf  Search for {\boldmath $B^- \to D_S^{(*)-} \phi$}
\end{center}
\bigskip

\begin{center}
\large The \babar\ Collaboration\\
\mbox{ }\\
\today
\end{center}
\bigskip \bigskip

\begin{center}
\large \bf Abstract
\end{center}

We report on searches for $B^- \to D_S^- \phi$ and 
$B^- \to D_S^{*-} \phi$.  In the context of the Standard Model 
the branching fractions for these decays are expected to
be highly suppressed, since they
proceed through annhilation of the $b$ and $\bar{u}$ quarks
in the $B^-$ meson.  Our results are based on 234 million
$\Upsilon{(42)} \to B\kern 0.18em\overline{\kern -0.18em B}$ decays collected with the 
\mbox{\slshape B\kern-0.1em{\smaller A}\kern-0.1em B\kern-0.1em{\smaller A\kern-0.2em R}}
 detector at SLAC.  We find no evidence for
these decays, and we set 90\% confidence level upper
limits on the branching fractions 
$${\cal B}(B^- \to D_S^- \phi)<1.8 \times 10^{-6}$$ 
$${\cal B}(B^- \to D_S^{*-} \phi)<1.1 \times 10^{-5}.$$
These results are consistent with Standard Model expectations.

\vfill

\begin{center}
Contributed to the 
XXII$^{\rm st}$ International Symposium on Lepton and Photon Interactions at High~Energies, 6/27 --- 7/5/2005, Uppsala, Sweden

\end{center}

\vspace{1.0cm}
\begin{center}
{\em Stanford Linear Accelerator Center, Stanford University, 
Stanford, CA 94309} \\ \vspace{0.1cm}\hrule\vspace{0.1cm}
Work supported in part by Department of Energy contract DE-AC03-76SF00515.
\end{center}

\newpage
} 

\begin{center}
\small

The \babar\ Collaboration,
\bigskip

B.~Aubert,
R.~Barate,
D.~Boutigny,
F.~Couderc,
Y.~Karyotakis,
J.~P.~Lees,
V.~Poireau,
V.~Tisserand,
A.~Zghiche
\inst{Laboratoire de Physique des Particules, F-74941 Annecy-le-Vieux, France }
E.~Grauges
\inst{IFAE, Universitat Autonoma de Barcelona, E-08193 Bellaterra, Barcelona, Spain }
% ...

\end{center}\newpage

\setcounter{footnote}{0}


\section{Introduction}
\label{sec:Introduction}
\begin{figure}[htb]
\begin{center}
\epsfig{file=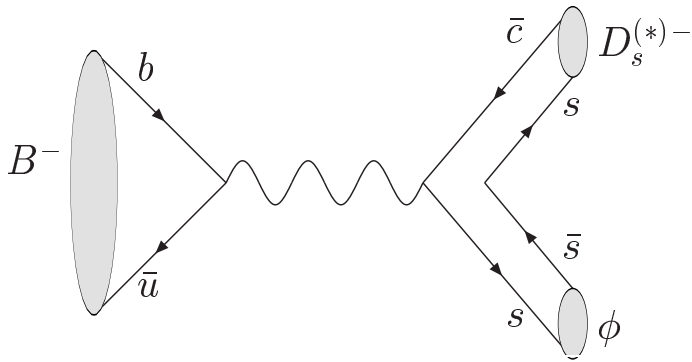,height=5cm}
\caption{Feynman diagram for $B^- \to D_S^{(*)-} \phi$.}
\label{fig:feyn}
\end{center}
\end{figure}

In the Standard Model (SM), the decay $B^- \to D_S^{(*)-} \phi$ 
occurs through annhilation of the two quarks in the
$B$-meson into a virtual $W$, see Figure~\ref{fig:feyn}.
No annhilation-type $B$ decays have ever been observed to date.
The current 90\% C.L. upper limits on $B^- \to D_S^- \phi$ and 
$B^- \to D^{*-}_S \phi$ are $3.2 \times 10^{-4}$ and 
$4 \times 10^{-4}$, respectively~\cite{cleo}.

In the SM, annhilation diagrams are highly suppressed.
Calculations of the $B^- \to D_S^{-} \phi$ 
branching fraction give predictions of $3 \times 10^{-7}$ 
using a perturbative QCD approach~\cite{lu}, or 
$7 \times 10^{-7}$ using QCD-improved factorization~\cite{mohanta}.

Since the 
current experimental limits are about three orders of magnitude 
higher than the SM expectations, searches for $B^- \to D_S^{(*)-} \phi$
could be sensitive to new physics contributions.  
For example, in Reference~\cite{mohanta} the branching fraction 
of $B^- \to D_S^{-} \phi$ is estimated to be $8 \times 10^{-6}$
in a two Higgs doublet model and $3 \times 10^{-4}$ 
in the minimal supersymmetric model
with $R$-parity violation.


\section{\boldmath The \babar\ detector and dataset}
\label{sec:babar}
Our results are based on $234 \times 10^6$ $\FourS\to B\Bbar$ decays,
corresponding to an integrated luminosity of 212 fb$^{-1}$, collected
between 1999 and 2004 with the \babar\ detector~\cite{babar} at the 
\pep2\  \BF\ at SLAC~\cite{pep2}.
A 12~fb$^{-1}$ off-resonance data sample,
with a center of mass (CM) energy 40~\mev below the \FourS resonance peak,
is used to study continuum events, $e^+ e^- \to q \bar{q}$
($q=u,d,s,$ or $c$).
The number of $B$-mesons in our data
sample is two orders of magnitudes larger than in the previously 
published search for $B^- \to D^{(*)-}_S \phi$~\cite{cleo}.


\section{\boldmath Analysis method}
\label{sec:Analysis}

We search for the decay
$B^- \to D^{(*)-}_S \phi$
in the following modes:
$D^{*-}_S \to D_S^- \gamma$,
$D_S^- \to \phi \pi^-$, $K_S K^-$, and $K^{*0}K^-$,
$\phi \to K^+ K^-$, $K_S \to \pi^+ \pi^-$, and
$K^{*0} \to K^+ \pi^-$
(charged conjugate decay modes are implied throughout this article).
We denote the $\phi$ from $B^- \to D_S^{(*)-} \phi$ decay
as the ``bachelor $\phi$'', in order to distinguish it
from the $\phi$ in the $D_S^- \to \phi \pi^-$ decay.

All kaon candidate tracks in the reconstructed decay chains must
satisfy a set of loose kaon identification criteria based on the
response of the internally-reflecting ring-imaging Cherenkov radiation
detector and the ionization measurements in the Drift Chamber
and the Silicon Vertex Tracker. The kaon
selection efficiency is a function of momentum and polar angle, and is
typically 95\%.  These requirements provide a rejection factor of
order 10 against pion backgrounds.  No particle identification
requirements are imposed on pion candidate tracks.

We select $\phi$, $K_S$, and $K^{*0}$ candidates from pairs of
oppositely-charged tracks with invariant masses consistent with the
parent particle decay hypothesis and consistent with originating from
a common vertex.  The invariant mass requirements are $\pm 10$ MeV
($\sim 2.4 \Gamma$) for the $\phi$, $\pm 9$ MeV ($\sim 3 \sigma$) for
the $K_S$, and $\pm 75$ MeV ($\sim 1.5 \Gamma$) for the $K^{*0}$.  We
then form $D_S^-$ candidates in the three modes listed above by
combining $\phi$, $K_S$, or $K^{*0}$ candidates with an additional
track.  The invariant mass of the $D_S^-$ candidate must be within 15
MeV ($\sim 3\sigma$) of the known $D_S^-$ mass.  In the $D_S^- \to
\phi \pi^-$ and $D_S^- \to K^{*0}K^-$ modes, all three charged tracks
are required to originate from a common vertex.  In the $D_S^- \to K_S
K^-$ mode, the $K_S$ and $D_S^-$ vertices are required to be separated
by at least 3 mm.  This last requirement is very effective in
rejecting combinatorial background and is 94\% efficient for signal.
We select $D^{*-}_S$ candidates from $D_S^-$ and photon candidates.
The photon candidates are constructed from calorimeter clusters with
lateral profiles consistent with photon showers and with energy above
60 MeV in the laboratory frame.  We require that the mass difference
$\Delta M$ between the $D^{*-}_S$ and $D_S$ candidates be between 130
and 156 MeV.  The $\Delta M$ resolution is about 5 MeV.

At each stage in the reconstruction chain, the measurement of the
momentum vector of an intermediate particle is improved by 
refitting the momenta of the decay products with kinematical
constraints.  These constraints are based on the known mass
of the intermediate particle and on the fact that the 
decay products must originate from a 
common point in space.

Finally, we select \Bm candidates by combining $D^{(*)-}_S$ and
bachelor $\phi$ candidates. A \Bm candidate is characterized
kinematically by the energy-substituted mass $\mes \equiv
\sqrt{(\frac{1}{2} s + \vec{p}_0\cdot \vec{p}_B)^2/E_0^2 - p_B^2}$ 
and energy difference $\Delta E \equiv E_B^*-\frac{1}{2}\sqrt{s}$,
where $E$ and $p$ are energy and momentum, the asterisk denotes the CM
frame, the subscripts $0$ and $B$ refer to the initial \FourS and $B$
candidate, respectively, and $s$ is the square of the CM energy.  In
the CM frame, $\mes$ reduces to $\mes = \sqrt{\frac{s}{4} -
p^{*2}_B}$.  For signal events we expect $\mes \sim M_B$, the known
$B^-$ mass, and $\Delta E \sim 0$.  The resolutions on
$\mes$ and $\Delta E$ are approximately 2.6 MeV and 10 MeV,
respectively.

If there is more than one $B^-$ candidate in an event, we retain the
best candidate based on a $\chi^2$ algorithm that uses the measured
values, known values, and resolutions for the $D_S^-$ mass, the
bachelor $\phi$ mass, and, where applicable, $\Delta M$.

Studies of simulated events and off-resonance data indicate that
most of the backgrounds to the $B^- \to D_S^{(*)-} \phi$ signal
are from continuum events.  To reduce these backgrounds
we make two additional requirements.  First, we require
$|\cos\theta_T| < 0.9$, where $\theta_T$ is the angle
between the thrust axes of the $B^-$ candidate and
the rest of the tracks and neutral clusters in the
event, calculated in the CM frame.  The distribution
of $|\cos\theta_T|$ is essentially uniform for signal
events and strongly peaked near one for continuum events.
Second, for each event we define a relative likelihood
for signal and background based on
a number of kinematical 
quantities.
The relative likelihood is defined as the ratio of the
likelihoods for signal and background.  The signal (background)
likelihood is defined as the product of the probability
density functions, PDFs, for the various kinematical
quantities in signal (background) events.  

The kinematical quantities used in the likelihood are 
reconstructed masses, decay angles, and a Fisher
discriminant designed to distinguish between 
continuum and $B\Bbar$ events.
All PDFs
are chosen based on studies of Monte Carlo and
off-resonance data.

The masses
used in the likelihoods are those of the $D_S^-$, the $\phi$ in the
$D_S^- \to \phi \pi^-$ , the $K^{*0}$ in
$D_S^- \to K^{*0} K^-$, and $\Delta M$ in 
$D_S^{*-} \to D_S \gamma$.  
The signal PDFs 
for the mass variables are the sum
of two Gaussian distributions for $D_S$ and $\Delta M$,
a Breit Wigner distribution for the $K^{*0}$, and
a Voigtian distribution for the $\phi$ from $D_S$ decay.  
The background
PDFs are constant functions.  Note that the
mass of the bachelor $\phi$ 
and the mass of the $K_S$ in $D_S^- \to K_S K^-$ 
are not used in the definition of the likelihoods.
This is because studies of background event samples
suggest that background events
contain mostly real bachelor $\phi$ and real $K_S$ mesons.

The decay angles used in the likelihood are those 
in the $K^{*0} \to K^+ \pi^-$ and in the $\phi \to K^+K^-$
decay, both for bachelor $\phi$ and the
$\phi$ from $D_S^- \to \phi \pi^-$ decay.
The signal PDFs for these quantities are set by angular
momentum conservation to be proportional to $\cos^2\theta$,
where $\theta$ is the decay angle for the process.
The one exception is the decay angle distribution of
the bachelor $\phi$ in $B^- \to D_S^{*-} \phi$ decay,
where the polarization of the two vector mesons in the
final state is not known.  For this reason, the decay
angle of the bachelor $\phi$ is not used in the definition
of the likelihood for the $B^- \to D_S^{*-} \phi$ mode.
The background PDFs for these variables are constant
functions of $\cos\theta$.
In addition, in the likelihood we also use the 
polar angle of the $B^-$ candidate in the CM frame ($\theta_B$).
The signal is expected to follow a $\sin^2\theta_B$ distribution,
while the background is independent of $\cos\theta_B$.

The final component of the likelihood is a Fisher discriminant
constructed from the quantities $L_0 = \sum_i{p_i}$ and 
$L_2 = \sum_i{p_i \cos^2\theta_i}$
calculated in the CM frame.  Here, $p_i$ is the momentum and
$\theta_i$ is the angle with respect to the thrust axis of the $B^-$ 
candidate
of tracks and clusters not used to reconstruct the $B^-$.
The signal and background PDFs for this variable are modelled
as bifurcated Gaussians with different means and standard deviations.
Note that this Fisher discriminant is highly correlated with the
$|\cos\theta_T|$ variable defined above.  It is because of this
correlation that the $|\cos\theta_T|$ variable is treated separately
and not included in the likelihood.

The combined efficiency of the requirements on likelihood and 
$|\cos\theta_T|$ varies between 71 and 83\%, depending
on the mode.  These requirements provide a rejection factor
of about 7 against backgrounds.  They were chosen from
studies of off-resonance data as well as 
simulated background and signal events.

After applying the requirements on relative likelihood and 
$|\cos\theta_T|$, we also demand that $\Delta E$ be within 
30 MeV ($\sim 3 \sigma$) of its expected mean value for signal events.
This mean value is determined from simulation, and varies
between $-3$ and $0$ MeV, depending on the mode.

\begin{table}[hbt]
\begin{center}
\caption{\protect
Efficiency ($\epsilon_i$), branching fractions (${\rm BR}_i$),
and products of efficiency and branching fractions
for the modes used in the
$B^- \to D_S^{(*)-} \phi$ search. The uncertainties on 
the $\epsilon_i$ and ${\rm BR}_i$ are discussed in the text.
Here $\rm{BR}_i$ is the product of branching fractions for
the secondary and tertiary decays in the $i$-th decay mode.}
\begin{tabular}{l c c c} \hline
Mode                   & $\epsilon_i$ 
& BR$_i$ & $\epsilon_i \cdot {\rm BR}_i$ \\ \hline
$B^- \rightarrow D_S^- \phi$,~~
$D_S^- \to \phi \pi^-$   & 0.192 & $11.6 \cdot 10^{-3}$ & $2.22 \cdot 10^{-3}$\\
$B^- \to D_S^- \phi$,~~
$D_S^- \to K^-  K_S$     & 0.177 & $8.20 \cdot 10^{-3}$ & $1.45 \cdot 10^{-3}$\\
$B^- \to D_S^- \phi$,~~
$D_S^- \to K^{*0} K^-$   & 0.140 & $14.5 \cdot 10^{-3}$ & $2.03 \cdot 10^{-3}$\\
\hline
$B^- \to D_S^{*-} \phi$,~~
$D_S^- \to \phi\pi^-$ & 0.109 & $10.9 \cdot 10^{-3}$ & $1.19 \cdot 10^{-3}$\\
$B^- \to D_S^{*-} \phi$,~~
$D_S^- \to K^-  K_S$     & 0.100 & $7.70 \cdot 10^{-3}$ & $0.77 \cdot 10^{-3}$\\
$B^- \to D_S^{*-} \phi$,~~
$D_S^- \to K^{*0} K^-$   & 0.083 & $13.6 \cdot 10^{-3}$ & $1.14 \cdot 10^{-3}$\\ 
\hline\hline
\end{tabular}
\label{tab:accBR}
\end{center}
\end{table}

The efficiencies of our selection requirements, shown in
Table~\ref{tab:accBR}, are determined from simulations.  In the case
of the $B^- \to D_S^{*-}\phi$ mode we take the average of the
efficiencies calculated assuming fully longitudinal or transverse
polarization for the two vector meson final state.
These efficiencies are found to be the same to within 1\%.  The
quantities BR$_i$ in Table~\ref{tab:accBR} are the product of the
known branching fractions for the secondary decay modes.  These are
taken from the compilation of the Particle Data Group~\cite{PDG}, with
the exception of the branching fraction for the $D_S \to \phi \pi$
mode, for which we use the latest most precise measurement ${\cal
B}(D_S \to \phi \pi) = (4.8 \pm 0.6)$\%~\cite{dsphipi}.  Since the
branching fractions for the other two $D_S$ modes are measured with
respect to the $D_S \to \phi \pi$ mode, we have also rescaled their
tabulated values from the Particle Data Group accordingly.


\section{\boldmath Systematic studies}
\label{sec:Systematics}

The systematic uncertainties on the products of efficiency and
branching ratio for the secondary decays in the decay chain of
interest are summarized in Table~\ref{tab:accErr}.  The largest
systematic uncertainty is associated with the uncertainty on the $D_S
\to \phi \pi$ branching ratio, which is only known to
12\%~\cite{dsphipi}, and which is used to normalize all other $D_S$
branching ratios.

From a purely experimental point of view, the most important
uncertainty is due to the uncertainty in the efficiency of the kaon
identification requirements.  The efficiency of these requirements is
calibrated using a sample of kinematically identified $D^{*0} \to D^0
\pi^+$, $D^{0} \to K^- \pi^+$, and is known at the level of 2\%.
Thus, this uncertainty result in a systematic uncertainty of 8\% for
the efficiency of the modes with four charged kaons, i.e. those with
$D^-_S \to K^{*0}K^-$ and $D_S^- \to \phi \pi$, and 6\% for the mode
with three charged kaons ($D_S^- \to K_S K^-$).  A second class of
uncertainties is associated with the detection efficiency for tracks
and clusters in the \babar\ detector.  From studies of a variety of
control samples, the tracking efficiency is understood at the level of
1.4\% or 0.6\% for transverse momenta below or above 200 MeV/c.  There
is also a 1.9\% uncertainty associated wih the reconstruction of the
$K_S \to \pi^+\pi^-$ decay which can occur a few centimeters away from
the interaction point.  Given the multiplicity and momentum spectrum
of tracks in the decay modes of interest, the uncertainty on the
efficiency to reconstruct the tracks in the $B$-decay chain is
estimated to be 3.7\%.  In the $B^- \to D_S^{*-} \phi$ search there is
an additional uncertainty of 1.8\% due to the uncertainty on the
efficiency to reconstruct the photon in the $D_S^{*-} \to D_S \gamma$
decay, and also a 1\% uncertainty from the unknown polarization in the
final state.  Finally, to ascertain the systematic due to the
efficiency of the other event selection requirements, we compute
the following efficiency variations: shifting the $\Delta E$ by
3\mev (0.3\%); shifting the mean of the $D_S$ and $\phi$ masses and
$\Delta M$ by 1\mev~ (0.2\%, 0.1\%, 0.2\%, respectively); increasing
the width of the $D_S$ and $\phi$ masses and $\Delta M$ by 1\mev~
(1.5\%, 0.4\%, 1.5\%, respectively); using a Fisher distribution
obtained from the data sample of a similar analysis, $B\to D\pi$ with
$D\to K\pi$ (3\%).  Thus we assign a 5\% systematic on the combined
efficiency of these selection criteria.

\begin{table}[hbt]
\begin{center}
\caption{\protect
Systematic uncertainties on $\sum_i \epsilon_i \cdot {\rm BR}_i$,
where the index $i$ runs over the three $D_S^-$ modes 
used in this analysis, $\epsilon_i$ are the experimental
efficiencies and ${\rm BR}_i$
are the branching fractions for the i-th mode.}
\begin{tabular}{l c c} \hline \hline
  & $B^- \to D_S^- \phi$ & $B^- \to D_S^{*-} \phi$ \\ \hline
$D_S$ branching fraction & 14\% & 14\% \\
$D_S^*$ branching fraction & - & 2.5\% \\
Other branching fractions & 1.5\% & 1.5\% \\
Charged kaon ID & 7.5\% & 7.5\% \\
Selection requirements & 5\% & 5\% \\ 
Tracking and $K_S$ efficiency & 3.7\% & 3.7\% \\
Photon efficiency & - & 1.8\% \\
Final state polarization & - & 1\% \\
Simulation statistics & 0.6\% & 0.6\% \\
\hline
Total & 17\% & 17\% \\ 
\hline\hline
\end{tabular}
\label{tab:accErr}
\end{center}
\end{table}


\section{\boldmath Physics results}
\label{sec:Physics}

We determine the yield of signal events from an unbinned extended
maximum-likelihood fit to the $\mes$ distribution of $B^-$ candidates
satisfying all of the requirements listed above.  
We fit simultaneously in two $|\Delta E|$ regions:
In the signal region the distribution is
parametrized as a Gaussian and the combinatorial
background as a threshold function~\cite{ARGUS}; in a sideband of
$\Delta E$ ($|\Delta E| < 200$ MeV, excluding the
signal region) we fit solely for the $\zeta$ parameter of the
threshold function.
In our fit, the
amplitude of the Gaussian is allowed to fluctuate to negative values,
but, for reasons of numerical stability, the sum of the Gaussian and
the threshold function is constrained to be positive over the full
$\mes$ fit range.  The mean and the standard deviations of the
Gaussian are constrained to the values determined from Monte Carlo
simulation. The fitting procedure was extensively tested with 
sets of simulated data, and was found to provide an unbiased estimate
of the signal yield.

\begin{figure}[hbt]
\begin{center}
\epsfig{file=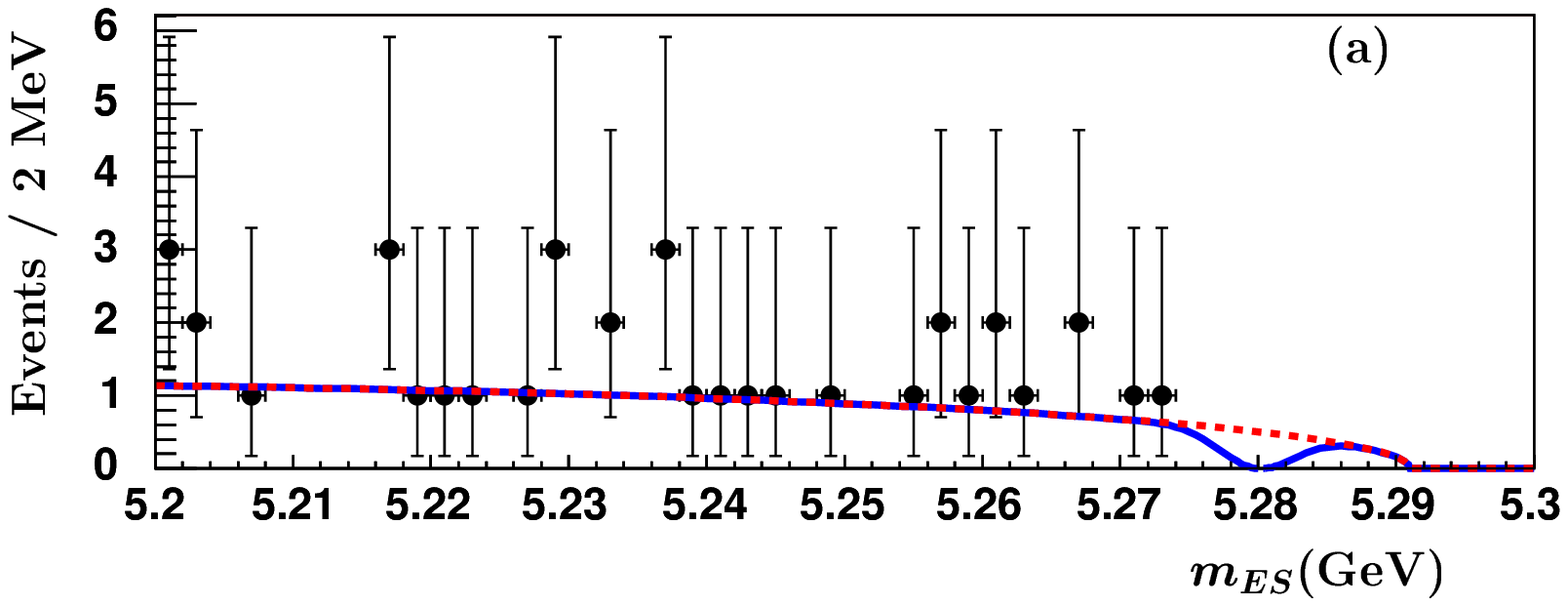,width=\linewidth}
\epsfig{file=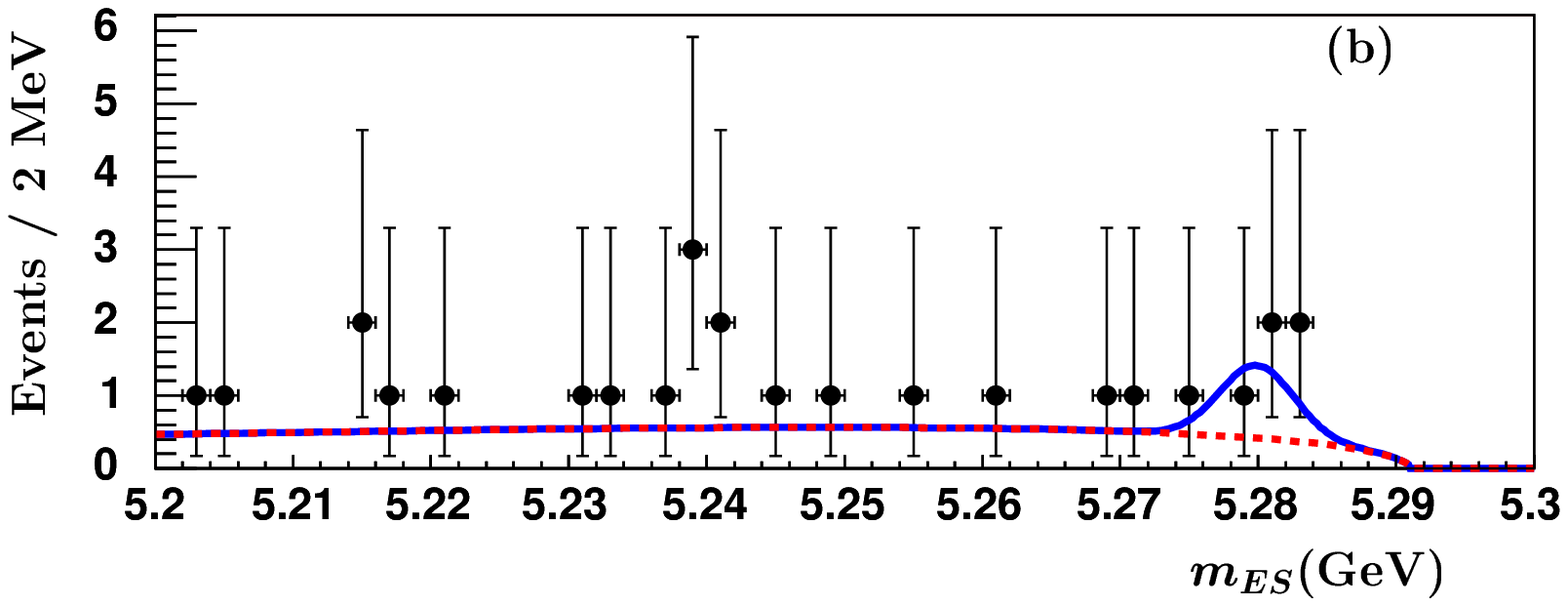,width=\linewidth}
\caption{Distribution of $\mes$ for (a) $B^- \to D_S^- \phi$ and
(b) $B^- \to D_S^{*-} \phi$ candidates.  The superimposed curves are
the result of the fits described in the text. The dashed red curve is the 
background contribution and the solid blue curve is the sum of the 
signal and backgound components.}
\label{fig:mesfit}
\end{center}
\end{figure}

\begin{figure}[hbt]
\begin{center}
\epsfig{file=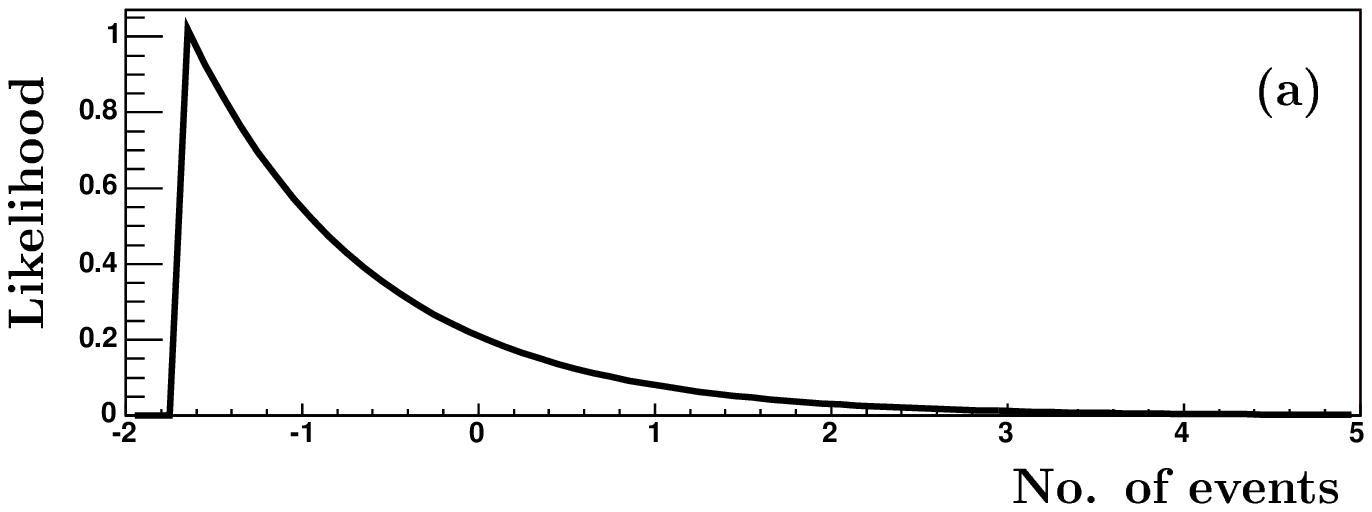,width=\linewidth}
\epsfig{file=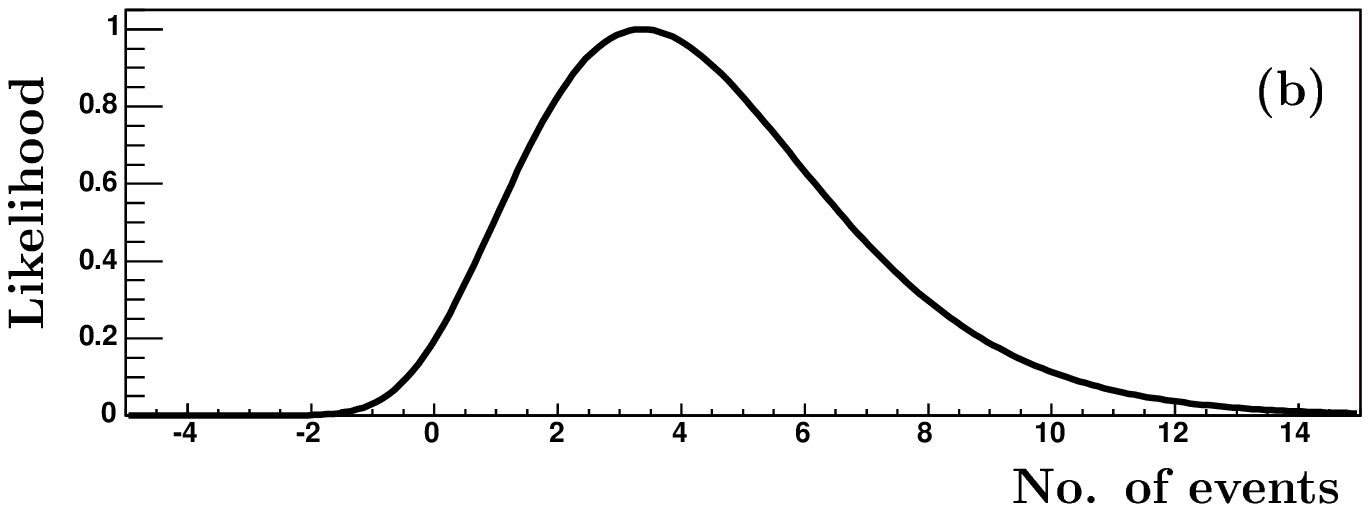,width=\linewidth}
\caption{Likelihood from the fit in arbitrary units as a 
function of the number of signal events.
(a) $B^- \to D_S^- \phi$;
(b) $B^- \to D_S^{*-} \phi$.}
\label{fig:lik}
\end{center}
\end{figure}

Figure~\ref{fig:mesfit} shows the $\mes$ distribution of the selected
candidates.  We see no evidence for $B^- \to D_S^{(*)-} \phi$.  The
fitted event yields are $N = -1.6^{+0.7}_{-0.0}$ and $N =
3.4^{+2.8}_{-2.1}$ for the $B^- \to D_S^- \phi$ and $B^- \to D_S^{*-}
\phi$ modes, respectively, where the quoted uncertainties correspond
to changes of $\frac{1}{2}$ in the log-likelihood for the fit.  The
likelihood curves are shown in Figure~\ref{fig:lik}.  The requirement
that the sum of the Gaussian and the threshold function be always
positive results in an effective constraint $N > -1.6$ in the $B \to
D_S \phi$ mode.  This is the source of the sharp edge at $N = -1.6$ in
the likelihood distribution of Figure~\ref{fig:lik}(a).

We use a Bayesian approach with a flat prior to set 90\% confidence
level upper limits on the branching fractions for the $B^- \to D_S^-
\phi$ and $B^- \to D_S^{*-} \phi$ modes.  In a given mode, the upper
limit on the number of observed events ($N_{UL}$) is defined as

\begin{equation}
\int_0^{N_{UL}} {\cal L}(N)~dN~~~=~~~\frac{9}{10} \int_0^{+\infty}{\cal L}(N)~dN\nonumber
\end{equation}

\noindent where ${\cal L}(N)$ is the likelihood as a function of the number
of signal events $N$ as determined from the $\mes$ fit described
above.  Then the upper limit ${\cal B}$ on the branching fraction is

\begin{equation}
{\cal B} < \frac{N_{UL}}{N_{BB} \sum_i \epsilon_i \cdot {\rm BR}_i} \nonumber
\nonumber
\end{equation}

\noindent where $N_{BB} = (233.9 \pm 2.5) \times 10^6$
is the number of $B\Bbar$ events, $i$ is an index that runs through
the three $D_S^-$ decay modes, $\epsilon_i$ is the acceptance in the
i-th mode, and BR$_{i}$ is the product of all secondary and
tertiary branching fractions (see Table~\ref{tab:accBR}).

We account for systematic uncertainties by numerically convolving
${\cal L}(N)$ with a Gaussian distribution with width determined by
the total systematic uncertainties (Table~\ref{tab:accErr}) in the two
modes, including the 1.1\% uncertainty in $N_{BB}$ added in
quadrature.  We find limits ${\cal B}(B^- \to D_S^- \phi) < 1.8 \times
10^{-6}$ and ${\cal B}(B^- \to D_S^{*-} \phi) < 1.1 \times 10^{-5}$ at
the 90\% confidence level.  As in Section~\ref{sec:Analysis}, these limits are
calculated using ${\cal B}(D_S \to \phi \pi) = (4.8 \pm 0.6)$\% from
Reference~\cite{dsphipi}.  
If we were to use the value ${\cal B}(D_S
\to \phi \pi) = (3.6 \pm 0.9)$\% from the Particle Data Group, we
would find ${\cal B}(B^- \to D_S^- \phi) < 2.6 \times 10^{-6}$ and
${\cal B}(B^- \to D_S^{*-} \phi) < 1.7 \times 10^{-5}$.  For
completeness, we also compute ${\cal B}(B^- \to D_S^- \phi) \times
{\cal B}(D_S \to \phi \pi) < 8.1 \times 10^{-8}$ and ${\cal B}(B^- \to
D_S^{*-} \phi) \times {\cal B}(D_S \to \phi \pi) < 5.3 \times
10^{-7}$, also at the 90\% confidence level.


In summary, we have searched for $B^- \to D_S^{(*)-} \phi$,
and we have found no evidence for these decays.
Our limits are about two
orders of magnitude lower than the previous results,
but are still one order of magnitude higher than the
Standard Model expectation.  Our limits, however, are 
much lower than expectations based on
$R$-parity violating supersymmetric models~\cite{mohanta}. 
The upper limit in the $B^- \to D_S^- \phi$ mode is also 
about a factor of four below the expectation from
a two Higgs doublet model~\cite{mohanta}.

\section{Acknowledgments}
\label{sec:Acknowledgments}
We are grateful for the 
extraordinary contributions of our \pep2\ colleagues in
achieving the excellent luminosity and machine conditions
that have made this work possible.
The success of this project also relies critically on the 
expertise and dedication of the computing organizations that 
support \babar.
The collaborating institutions wish to thank 
SLAC for its support and the kind hospitality extended to them. 
This work is supported by the
US Department of Energy
and National Science Foundation, the
Natural Sciences and Engineering Research Council (Canada),
Institute of High Energy Physics (China), the
Commissariat \`a l'Energie Atomique and
Institut National de Physique Nucl\'eaire et de Physique des Particules
(France), the
Bundesministerium f\"ur Bildung und Forschung and
Deutsche Forschungsgemeinschaft
(Germany), the
Istituto Nazionale di Fisica Nucleare (Italy),
the Foundation for Fundamental Research on Matter (The Netherlands),
the Research Council of Norway, the
Ministry of Science and Technology of the Russian Federation, and the
Particle Physics and Astronomy Research Council (United Kingdom). 
Individuals have received support from 
CONACyT (Mexico),
the A. P. Sloan Foundation, 
the Research Corporation,
and the Alexander von Humboldt Foundation.


\end{document}